\documentclass[twocolumn]{aastex631}
\usepackage[utf8]{inputenc}
\usepackage{natbib}
\usepackage{graphicx}
\usepackage{xcolor}
\usepackage{threeparttable}

\received{\today}
\revised{\today}
\accepted{\today}
\submitjournal{ApJL}

\shorttitle{LTT 9779\,{\rm b} NIRISS/SOSS Transmission}
\shortauthors{Radica et al.}

\begin{document}

\title{Muted Features in the JWST NIRISS Transmission Spectrum of Hot-Neptune LTT~9779~b}

\correspondingauthor{Michael Radica}
\email{michael.radica@umontreal.ca}


\author[0000-0002-3328-1203]{Michael Radica}
\affiliation{Institut Trottier de Recherche sur les Exoplanètes and Département de Physique, Université de Montréal, 1375 Avenue Thérèse-Lavoie-Roux, Montréal, QC, H2V 0B3, Canada}

\author[0000-0002-2195-735X]{Louis-Philippe Coulombe}
\affiliation{Institut Trottier de Recherche sur les Exoplanètes and Département de Physique, Université de Montréal, 1375 Avenue Thérèse-Lavoie-Roux, Montréal, QC, H2V 0B3, Canada}

\author[0000-0003-4844-9838]{Jake Taylor}
\affiliation{Department of Physics, University of Oxford, Parks Rd, Oxford OX1 3PU, UK}

\author[0000-0003-0475-9375]{Loic Albert}
\affiliation{Institut Trottier de Recherche sur les Exoplanètes and Département de Physique, Université de Montréal, 1375 Avenue Thérèse-Lavoie-Roux, Montréal, QC, H2V 0B3, Canada}

\author[0000-0002-1199-9759]{Romain Allart}
\affiliation{Institut Trottier de Recherche sur les Exoplanètes and Département de Physique, Université de Montréal, 1375 Avenue Thérèse-Lavoie-Roux, Montréal, QC, H2V 0B3, Canada}

\author[0000-0001-5578-1498]{Björn Benneke}
\affiliation{Institut Trottier de Recherche sur les Exoplanètes and Département de Physique, Université de Montréal, 1375 Avenue Thérèse-Lavoie-Roux, Montréal, QC, H2V 0B3, Canada}

\author[0000-0001-6129-5699]{Nicolas B.\ Cowan}
\affiliation{Department of Physics, McGill University, 3600 rue University, Montréal, QC, H3A 2T8, Canada}
\affiliation{Department of Earth and Planetary Sciences, McGill University, 3600 rue University, Montréal, QC, H3A 2T8, Canada}

\author[0000-0003-4987-6591]{Lisa Dang}
\altaffiliation{Banting Postdoctoral Fellow}
\affiliation{Institut Trottier de Recherche sur les Exoplanètes and Département de Physique, Université de Montréal, 1375 Avenue Thérèse-Lavoie-Roux, Montréal, QC, H2V 0B3, Canada}

\author[0000-0002-6780-4252]{David Lafrenière}
\affiliation{Institut Trottier de Recherche sur les Exoplanètes and Département de Physique, Université de Montréal, 1375 Avenue Thérèse-Lavoie-Roux, Montréal, QC, H2V 0B3, Canada}

\author[0000-0002-5113-8558]{Daniel Thorngren}
\affiliation{Department of Physics \& Astronomy, Johns Hopkins University, Baltimore, MD 21210, USA}
\altaffiliation{Davis Postdoctoral Fellow}

\author[0000-0003-3506-5667]{\'Etienne Artigau}
\affiliation{Institut Trottier de Recherche sur les Exoplanètes and Département de Physique, Université de Montréal, 1375 Avenue Thérèse-Lavoie-Roux, Montréal, QC, H2V 0B3, Canada}

\author[0000-0001-5485-4675]{René Doyon}
\affiliation{Institut Trottier de Recherche sur les Exoplanètes and Département de Physique, Université de Montréal, 1375 Avenue Thérèse-Lavoie-Roux, Montréal, QC, H2V 0B3, Canada}

\author[0000-0001-6362-0571]{Laura Flagg}
\affiliation{Department of Astronomy and Carl Sagan Institute, Cornell University, Ithaca, New York 14853, USA}

\author[0000-0002-6773-459X]{Doug Johnstone}
\affiliation{NRC Herzberg Astronomy and Astrophysics, 5071 West Saanich Rd, Victoria, BC, V9E 2E7, Canada}
\affiliation{Department of Physics and Astronomy, University of Victoria, Victoria, BC, V8P 5C2, Canada}

\author[0000-0002-8573-805X]{Stefan Pelletier}
\affiliation{Institut Trottier de Recherche sur les Exoplanètes and Département de Physique, Université de Montréal, 1375 Avenue Thérèse-Lavoie-Roux, Montréal, QC, H2V 0B3, Canada}

\author[0000-0001-6809-3520]{Pierre-Alexis Roy}
\affiliation{Institut Trottier de Recherche sur les Exoplanètes and Département de Physique, Université de Montréal, 1375 Avenue Thérèse-Lavoie-Roux, Montréal, QC, H2V 0B3, Canada}

\begin{abstract}
The hot-Neptune desert is one of the most sparsely populated regions of the exoplanet parameter space, and atmosphere observations of its few residents can provide insights into how such planets have managed to survive in such an inhospitable environment. Here, we present transmission observations of LTT~9779~b, the only known hot-Neptune to have retained a significant H/He-dominated atmosphere, taken with JWST NIRISS/SOSS. The 0.6 -- 2.85\,µm transmission spectrum shows evidence for muted spectral features, rejecting a perfectly flat line at $>$5$\sigma$. We explore water and methane-dominated atmosphere scenarios for LTT~9779~b's terminator, and retrieval analyses reveal a continuum of potential combinations of metallicity and cloudiness. Through comparisons to previous population synthesis works and our own interior structure modelling, we are able to constrain LTT~9779~b's atmosphere metallicity to 20--850$\times$ solar. Within this range of metallicity, our retrieval analyses prefer solutions with clouds at mbar pressures, regardless of whether the atmosphere is water- or methane-dominated --- though cloud-free atmospheres with metallicities $>$500$\times$ solar cannot be entirely ruled out. By comparing self-consistent atmosphere temperature profiles with cloud condensation curves, we find that silicate clouds can readily condense in the terminator region of LTT~9779~b. Advection of these clouds onto the day-side could explain the high day-side albedo previously inferred for this planet and be part of a feedback loop aiding the survival of LTT~9779~b's atmosphere in the hot-Neptune desert. 
\end{abstract}


\section{Introduction} 
\label{sec: Introduction}

\begin{figure} 
	\centering
	\includegraphics[width=\columnwidth]{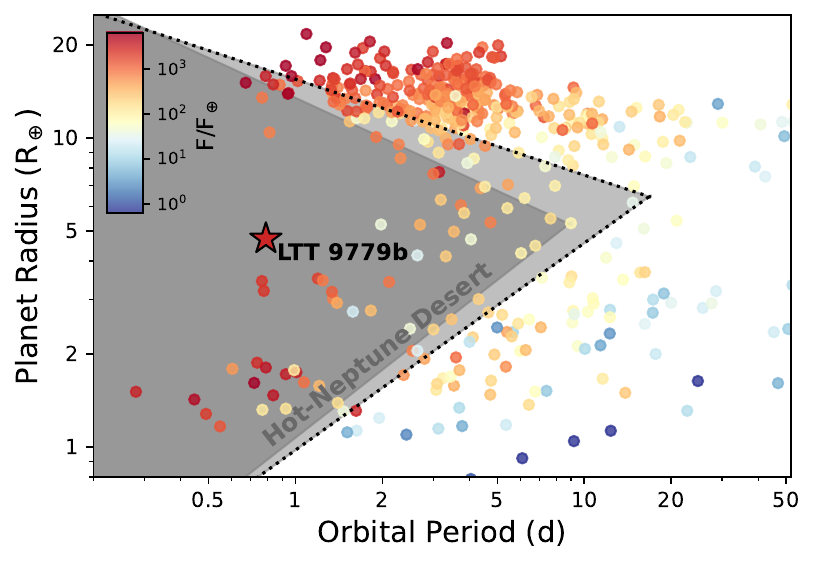}
    \caption{The population of close-in exoplanets visualized in period-radius space. All confirmed planets (from the NASA Exoplanet Archive), with masses and radii measured to better than 15\% and 10\% respectively, are shown coloured by the amount of flux they receive from their host star relative to the Earth. The hot-Neptune desert, as defined by \citet{mazeh_dearth_2016}, is marked in grey, and LTT~9779~b is identified with a star.
    \label{fig:NeptuneDesert}}
\end{figure}

In the decades since the discovery of the first transiting exoplanet around a Sun-like star \citep{charbonneau_detection_2000}, transit and radial velocity measurements have revealed the parameter space of close-in exoplanets \citep{borucki_kepler_2010, hartman_hat-p-39bhat-p-41b_2012, hebrard_wasp-52b_2013, hellier_transiting_2014, delrez_wasp-121_2016, cloutier_characterization_2017, radica_revisiting_2022}. Population-level studies of exoplanets have revealed several peculiar demographic features; one of particular interest being the hot-Neptune desert \citep[][Fig.~\ref{fig:NeptuneDesert}]{szabo_short-period_2011, mazeh_dearth_2016} --- a dearth of Neptune-mass planets with orbital periods shorter than $\sim$4\,d, which cannot be explained by observational biases. The hot-Neptune desert is predicted to be carved out by photoevaporation driven by high energy radiation from the host star \citep{owen_kepler_2013, owen_photoevaporation_2018}. Jupiter-mass planets are massive enough to retain the bulk of their atmosphere against photoevaporation, whereas smaller planets lose their entire gaseous envelope on short timescales \citep[$<$100\,Myr;][]{owen_kepler_2013} leaving behind bare rocky cores, with perhaps trace amounts of an atmosphere \citep[e.g.,][]{armstrong_remnant_2020}. Indeed, studies of atmosphere escape, via e.g., the 1.083\,µm metastable He triplet \citep{seager_theoretical_2000, oklopcic_new_2018}, have demonstrated that many of the planets bordering the hot-Neptune desert are currently experiencing substantial mass loss \citep{spake_helium_2018, allart_spectrally_2018, allart_homogeneous_2023}.

LTT~9779~b \citep{jenkins_ultra-hot_2020} is one of only a handful of known hot-Neptunes. With a mass of 29.32$\pm$0.8\,M$_\oplus$, a radius of 4.72$\pm$0.23\,R$_\oplus$, and a 0.792\,d period, it resides right in the heart of the hot-Neptune desert \citep[Fig.~\ref{fig:NeptuneDesert}, ][]{jenkins_ultra-hot_2020}. Despite orbiting in this inhospitable region of parameter space, observations with the  Spitzer Space Telescope's Infrared Array Camera (IRAC) provided evidence for a substantial H/He-rich atmosphere, with a metallicity $>$30$\times$ solar \citep{dragomir_spitzer_2020, crossfield_phase_2020}. \textit{Spitzer}/IRAC eclipse observations also revealed a day-side brightness temperature of $>$2300\,K at 3.6\,µm --- equivalent to some of the hottest hot-Jupiters \citep{hartman_hat-p-39bhat-p-41b_2012, delrez_wasp-121_2016, mansfield_unique_2021}. 

Recently, \citet{hoyer_extremely_2023} presented the detection of the optical eclipse of LTT~9779~b with a depth of 115$\pm$24\,ppm using the Charaterising Exoplanets Satellite (CHEOPS), indicating an extremely high ($\sim$0.8) day-side geometric albedo. Their 1D self-consistent models suggest that the high optical albedo is caused by silicate clouds on the day-side, which due to the planet's extreme day-side temperature, can only be possible if the atmospheric metallicity is $\gtrsim$400$\times$ solar \citep{hoyer_extremely_2023}. An analysis of follow-up observations with the UVIS mode of the Hubble Space Telescope (HST) is also underway to confirm this high albedo in the near-ultraviolet\footnote{GO 16915; PI Radica}. However, transmission observations with \textit{HST}'s Wide Field Camera 3 (WFC3) G102 and G141 grisms presented by \citet{edwards_characterizing_2023} suggest a cloud-free terminator and a very low (subsolar) atmospheric metallicity. They also find no evidence for atmospheric escape via an analysis of the He 1.083\,µm metastable triplet. 

Finally, \citet{Fernandez2023} recently published an upper limit on the XUV flux of LTT 9779 using XMM-Newton. They indicate that the X-ray faintness of the G7V host star likely played a major role in the long-term survival of LTT 9779~b's atmosphere.   

Here we present a transmission spectrum of LTT~9779~b with the Single Object Slitless Spectroscopy (SOSS) mode \citep{albert_near_2023} of JWST's Near Infrared Imager and Slitless Spectrograph (NIRISS) instrument \citep{doyon_near_2023}. This is the first paper in a three-part series: Coulombe et al.~(submitted) presents a complementary analysis of the full NIRISS/SOSS phase curve of LTT~9779~b, and Radica et al.~(in prep) will follow up with an in-depth analysis of the secondary eclipse spectrum. This manuscript is organized as follows: we briefly outline the observations and data reduction in Section~\ref{sec: Observations}, and the light curve fitting in Section~\ref{sec: Fitting}. We then describe our atmosphere modelling in Section~\ref{sec: Modelling}, follow up with a discussion in Section~\ref{sec: Discussion}, and concluding remarks in Section~\ref{sec: Conclusions}.

\section{Observations and Data Reduction} 
\label{sec: Observations}

We observed a full-orbit phase curve of LTT~9779~b with NIRISS/SOSS as part of GTO 1201 (PI: Lafrenière), starting on Jul 7, 2022 at 01:29:06\,UTC and lasting 21.9 hours. We operated SOSS in the SUBSTRIP256 readout mode, which provides access to the second diffraction order containing wavelengths shorter than $\sim$0.85\,µm \citep{albert_near_2023}. We thus obtained the full spectral coverage of SOSS, from 0.6 -- 2.85\,µm. The time series observations (TSO) used ngroup=2 and consisted of 4790 total integrations.

The data reduction and validation are treated more fully in Coulombe et al.~(submitted), but for completeness, the pertinent points are summarized below. We reduced the full phase curve using the \texttt{supreme-SPOON} pipeline \citep{radica_awesome_2023, feinstein_early_2023, lim_atmospheric_2023, fournier-tondreau_near-infrared_2023}, following Stages 1 -- 3  as laid out in \citet{radica_awesome_2023}. We extracted the 1D stellar spectra using a simple box aperture since the effects of the SOSS self-contamination are predicted to be negligible for this target \citep{darveau-bernier_atoca_2022, radica_applesoss_2022}. We used an extraction width of 40 pixels, which we found to minimize the out-of-transit scatter in the extracted spectra. 

In order to ensure that our results are robust against the particularities of a given reduction pipeline, we also reduced the full TSO with the NAMELESS pipeline \citep{feinstein_early_2023, coulombe_broadband_2023}. We performed steps of Stages 1 and 2 as described in \citet{coulombe_broadband_2023}, putting special emphasis on the correction of cosmic rays and 1/$f$ as we found those two effects to be the main contributors of scatter in the extracted spectra. Cosmic rays were detected and clipped using a running median of the individual pixel counts (see Coulombe et al., submitted for a more in-depth description). As for the 1/$f$ noise, we used the method of \citet{coulombe_broadband_2023} and considered only pixels less than 30 pixels away from the trace to compute the 1/$f$, similar to what is done in \citet{feinstein_early_2023} and \citet{Holmberg_2023}, as we found this method to significantly reduce scatter at longer wavelengths.

\section{Light Curve Fitting} 
\label{sec: Fitting}

In this work, we focus solely on the transit portion of the phase curve observations. We compare two methods of extracting the transmission spectrum from the observations: Method 1 uses the \texttt{supreme-SPOON} reduction, and fits light curves from the portion of the phase curve surrounding the transit. Method 2 uses the NAMELESS reduction and simultaneously fits the entire phase curve. 

\subsection{Method 1: Transit Fitting}
Using the \texttt{supreme-SPOON} reduction, we extract from the phase curve the integrations containing the transit of LTT~9779~b (integrations 2000 -- 3000 or BJD 2459767.44667 -- 2459767.63738). We opt for a slightly longer ($\sim$3.5:1) baseline in order to better constrain high-frequency variations in the TSO, which we attribute to stellar granulation (Coulombe et al.~submitted). In total, our transit observations last 4.57 hours, with 2 hours before and 1.7 hours after the $\sim$47-minute transit.

\subsubsection{White Light Curve Fitting}
\label{sec: wlc}

\begin{figure*} 
	\centering
	\includegraphics[width=\textwidth]{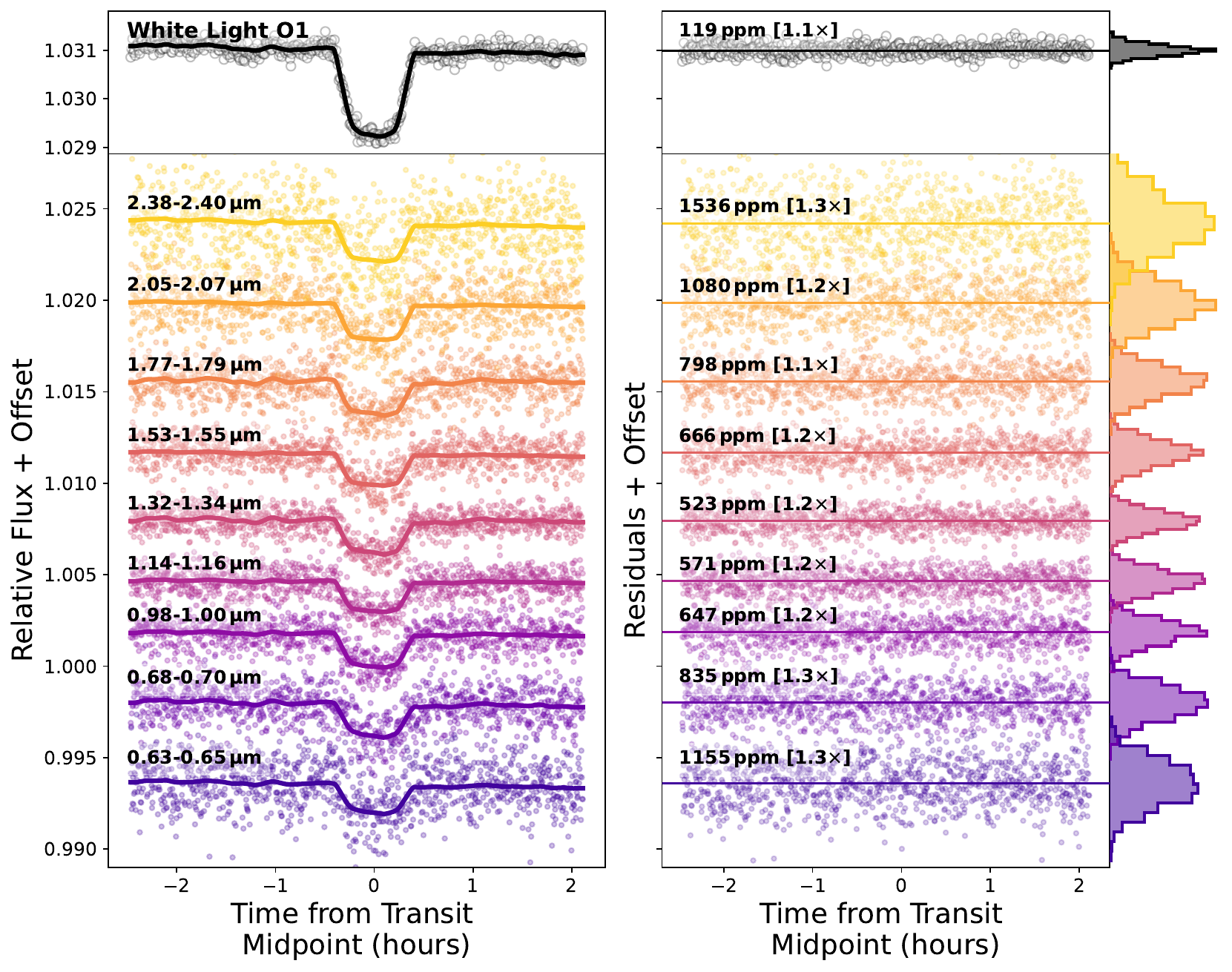}
    \caption{LTT~9779~b transit light curve fits. \emph{Left}: Order 1 white light curve, and examples of 10 spectrophotometric transit light curves (open points) and their best-fitting transit models (solid lines) from the \texttt{supreme-SPOON} reduction. The wavelength extent of each bin is shown above the best-fitting model.
    \emph{Right}: Residuals to each best-fitting model. The scatter in the residuals in ppm is noted above each, as well as, in brackets, its multiple above the predicted photon noise level. Histograms of the residuals are shown to the right.
    \label{fig:spec}}
\end{figure*}

We construct white light curves for each of the two SOSS orders by summing the flux from all wavelengths in order 1 and wavelengths $<$0.85\,µm in order 2. We jointly fit two transit models to the white light curves. Each transit model consists of an astrophysical model and a ``systematics" model. The astrophysical model is a standard \texttt{batman} transit model \citep{kreidberg_batman_2015}. We fix the orbital period to 0.79207022\,d, and assume a circular orbit \citep{crossfield_phase_2020}. The planet's orbital parameters, that is the time of mid-transit, $T_0$, the impact parameter, $b$, and the scaled semi-major axis, $a/R_*$ are shared between both orders, whereas we individually fit the transit radius, $R_p/R_*$, as well as the two parameters of the quadratic limb-darkening law, $u_1$ and $u_2$, to each order.

The ``systematics" model takes care of everything that is not the planetary transit. Low-frequency, non-Gaussian noise is clearly visible in the out-of-transit baseline. Coulombe et al.~(submitted) conclude that these deviations are likely caused by granulation in the host star's photosphere \citep[e.g.,][]{kallinger_connection_2014, grunblatt_seeing_2017, pereira_gaussian_2019}, and correct its effects using a Gaussian Process (GP) with simple harmonic oscillator (SHO) kernel. We follow the same prescription here. Using \texttt{celerite} \citep{foreman-mackey_fast_2017} we fit the two parameters of a SHO GP, $a_{gran}$ and $b_{gran}$ --- corresponding to the amplitude and characteristic timescale of the granulation signal \citep{pereira_gaussian_2019}. The timescale is shared between both orders, however, the GP amplitude is individually fit. Furthermore, for each order we fit for the transit baseline zero-point, as well as a linear trend to account for the planet's phase variation, and the long-term slope seen in the full phase curve data (Coulombe et al.~submitted). Lastly, for each order, we fit an error inflation term added in quadrature to the extracted errors. Our final transit model therefore has 18 parameters: nine for the astrophysical model, and another nine to account for ``systematics". We fit the light curves with \texttt{juliet}'s \citep{espinoza_juliet_2019} implementation of \texttt{pymultinest} \citep{buchner_statistical_2016}, using wide, uninformative priors for all parameters. The white light curve and best fitting model are shown in the top panel of Figure~\ref{fig:spec}, and relevant fitted parameters are summarized in Table~\ref{tab: WLC Parameters}. 

\begin{table}
\centering
\caption{Best-fitting white light transit parameters}
\label{tab: WLC Parameters}
\begin{threeparttable}
    \begin{tabular}{cc}
        \hline
        \hline
        Parameter & Value \\
        \hline
        T0 [BJD] & 59767.54935$\pm$1.5$\times$10$^{-4}$ \\
        $\rm R_p/R_{*,O1}$ & 0.04749$\pm$0.00261 \\
        $\rm R_p/R_{*,O2}$ & 0.04623$\pm$0.00308 \\
        $\rm b$ & 0.91687$\pm$0.00903 \\
        $\rm a/R_*$ & 3.80948$\pm$0.13473 \\
        $\rm a_{gran,O1}$ [ppm] & 68.66$\pm$20.04 \\
        $\rm a_{gran,O2}$ [ppm] & 95.73$\pm$29.35 \\
        $\rm b_{gran}$ [min] & 24.97$\pm$8.08 \\
        \hline
    \end{tabular}
    \begin{tablenotes}
        \small
        \item \textbf{Notes}: Subscripts O1 and O2 refer to values from the order 1 and order 2 white light curves respectively. Parameters without an order subscript were jointly fit to both orders.
    \end{tablenotes}
\end{threeparttable}
\end{table}

\subsubsection{Spectrophotometric Light Curve Fitting}
\label{sec: spec fitting}

We create spectrophotometric light curves by binning the data to a constant resolution of $R$=150. We choose $R$=150 as it results in a sufficient signal-to-noise in the binned light curves to accurately model the aforementioned systematics, though we note that our results remain unchanged if we instead fit at a higher resolution. Furthermore, binning the light curve before fitting results in more accurate transit depth precisions compared to a fitting-then-binning scheme, which inherently assumes that individual bins are uncorrelated, which, particularly for small bins, may not be the case \citep{Holmberg2023}.

We fix the orbital parameters in the spectrophotometric fits to those obtained from the transit fits (Table~\ref{tab: WLC Parameters}), and the quadratic limb-darkening parameters to model predictions from \texttt{ExoTiC-LD} \citep{david_grant_2022_7437681} using the 3D Stagger grid \citep{magic2015}, as the planet's high inclination ($\sim$76$\rm ^o$) makes it impossible to accurately back-out the stellar limb-darkening from the transit light curves \citep{muller_high-precision_2013}. When freely fit, the limb-darkening remains entirely unconstrained across the full prior range and a spurious upward slope towards redder wavelengths is introduced into the transmission spectrum. We also note that our results presented below are unchanged if we instead use other (i.e., four-parameter) limb-darkening law. We scale the best-fitting GP model from the white light curve to account for the granulation signal in each of the spectrophotometric light curves, as we find that the granulation noise only varies in amplitude with wavelength (becoming less prominent towards redder wavelengths) and not its overall structure. We also fit for the transit zero point and a linear trend. Our spectrophotometric light curve model therefore consists of five parameters. Examples of several bins and their best-fitting models are shown in Figure~\ref{fig:spec}.

\subsection{Method 2: Phase Curve Fitting}
Once again, the phase curve fitting methodology is outlined in further detail in Coulombe et al.~(submitted), but we provide a short summary here. Given the extreme equilibrium temperature of LTT~9779~b, it is possible that the ``dark planet'' assumption inherent in our transit-only fits breaks down and our retrieved transit depths are biased by non-negligible night side flux \citep{Kipping2010}. However, by fitting the full phase curve, we naturally account for the potential for the planet to have non-zero night side flux \citep[e.g.,][]{Martin-Lagarde2020}.

The phase curve fits were performed considering a 12-parameter slice model, with the planetary surface being divided into six longitudinal slices for which we each fit a thermal emission and albedo value, describing the variation of the planetary flux throughout its orbit \citep[e.g.,][]{Cowan2011}. For the white light fit, instead of using the two SOSS orders as was done with the transit-only analysis above, we constructed a phase curve dominated by reflected-light emission, using wavelengths $<$1\,µm, such that we needed only to consider albedo, in addition to a systematics model consisting of a SHO GP and linear slope as described in Section~\ref{sec: wlc}, at the white-light curve stage. This choice was made in order to enable the best-possible correction of the granulation signal, as we found that fitting for the albedo and thermal emission in addition to the SHO GP led to the phase curve model absorbing some of the granulation signal; resulting in nonphysical solutions.

Both thermal emission and reflected light were then accounted for in the fits for each spectrophotometric bin. To extract a transmission spectrum, we fit for the planetary radius and limb-darkening coefficients at each wavelength. The orbital parameters were fixed to the values from Table~\ref{tab: WLC Parameters}, and the limb-darkening coefficients to the same values considered in the transit-only analysis for easier comparison between the two reductions. Finally, for each bin we considered a linear slope as well as the scaled best-fitting white light curve GP for our spectroscopic systematics model. 

\begin{figure*} 
	\centering
	\includegraphics[width=\textwidth]{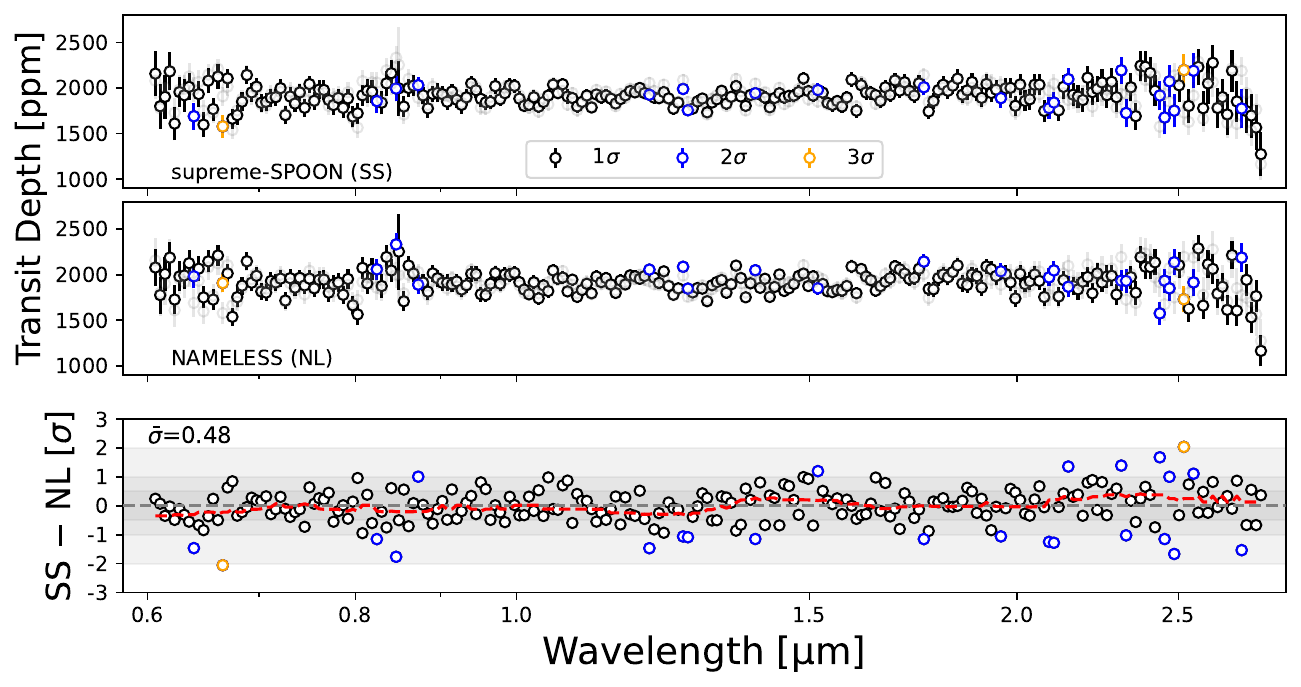}
    \caption{Comparison between \texttt{supreme-SPOON} and NAMELESS transmission spectra. 
    \emph{Top}: \texttt{supreme-SPOON} transmission spectrum with points colored by their deviation from the NAMELESS spectrum. Black points are consistent between the two spectra at a 1$\sigma$ level, blue points at 2$\sigma$, etc. Sigma levels are calculated using the maximum error bar between the two spectra for each wavelength bin. The NAMELESS spectrum is also shown in faded points.
    \emph{Middle}: Same as the top panel, but for the NAMELESS spectrum.
    \emph{Bottom}: Differences in each spectral bin divided by the maximum error bar of the two reductions. The red dashed line is a running median showing that there are no low-frequency trends. The average deviation between the two spectra is 0.48$\sigma$. 
    \label{fig:spec_compare}}
\end{figure*}

\subsection{Spectrum Consistency}
There is good agreement between the \texttt{supreme-SPOON} and NAMELESS transmission spectra, which have not only been reduced with independent pipelines but also fit with different methodologies (Figure~\ref{fig:spec_compare}). The spectra are consistent with each other at a 0.48$\sigma$ level, on average, with no systematic biases. Nevertheless, we recognize that even small deviations, or single outliers between spectra that look consistent by-eye can lead retrievals to divergent inferences \citep[e.g.,][]{Welbanks2023}. We thus conduct our analysis on a combination of our two transmission spectra to pseudo-marginalize over any differences in data reduction and light curve fitting techniques \citep[e.g.,][]{bell_methane_2023}. We thus create a combined transmission spectrum by taking the average of the transit depths in each wavelength bin. We then assign the transit depth error as the maximum of the errors from each spectrum in the case that the points are 1$\sigma$ consistent, or calculate a new error bar by inflating the maximum error until 1$\sigma$ consistency is achieved. In the remainder of this study, we use the combined transmission spectrum, though we note that our results are consistent if we focus on either reduction alone.

\section{Atmosphere Modelling} 
\label{sec: Modelling}

To explore the plausible range of atmospheric scenarios that are consistent with the data, we used the open-source atmospheric retrieval tool CHIMERA \citep{Line2013, taylor_awesome_2023}. We start with a free chemistry retrieval considering the following species which are commonly expected in the atmospheres of transiting planets: H$_2$O \citep{Polyansky2018, Freedman2014}, CH$_4$ \citep{Rothman+10}, CO \citep{Rothman+10}, CO$_2$ \citep{Freedman2014}, Na \citep{Kramida2018, Allard2019} and K \citep{Kramida2018, Allard2016}, and limit the vertically-constant log abundances of each species between [-12, -1], except for H$_2$O and CH$_4$ for which we use [-12, 0] in order to probe extremely high metallicity scenarios which cannot be ruled out for Neptune-mass planets \citep[e.g.,][]{fortney_framework_2013, moses_compositional_2013, morley_forward_2017}. For H/He-dominated atmospheres, we include H$_2$-H$_2$ and H$_2$-He collision-induced absorption \citep{Richard2012}. The isothermal atmosphere temperature can vary between 700--2000\,K, based on the range of temperature profiles around the terminator region retrieved by Coulombe et al.~(submitted). We also include a grey cloud deck, constrained to pressures between 1 and 10$^{-6}$\,bar. 

We do not consider the transit light source effect \citep[TLSE;][]{Rackham2018, lim_atmospheric_2023, fournier-tondreau_near-infrared_2023} in our retrievals as LTT~9779 is known to be slowly rotating and thus inactive \citep{jenkins_ultra-hot_2020}. Moreover, \citet{edwards_characterizing_2023} included the TLSE in their HST retrievals but found that it was disfavoured over models without the TLSE, and resulted in unreasonably high spot and faculae filling factors. 

Unsurprisingly, given the muted nature of the spectrum, there are no strong constraints on the atmosphere composition. The retrieval finds evidence for some features due to a combination of H$_2$O and CH$_4$, as well as a preference for clouds at $\sim$mbar pressure levels; though constraints on each of these parameters are weak (e.g., Figure~\ref{fig:corner}). The abundances of all other species considered in the retrieval remain unconstrained by the data. 

Within the SOSS waveband, transmission observations are primarily sensitive to absorption due to H$_2$O absorption bands at 1.2, 1.4, 1.8\,µm, etc.~\citep{feinstein_early_2023, radica_awesome_2023, taylor_awesome_2023}. Depending on the thermochemical regime of the atmosphere (T$\lesssim$850\,K or C/O$>$1), absorption bands of CH$_4$ could potentially be visible at 1.4, 1.7, 2.3\,µm \citep{moses_disequilibrium_2011, visscher_quenching_2011, madhusudhan_co_2012}. Other major C-bearing molecular species (e.g., CO, CO$_2$) have negligible opacity in the SOSS waveband and we cannot reasonably expect to constrain them with SOSS observations alone \citep{taylor_awesome_2023}. 

Based on the above considerations, we conduct a series of nested retrievals to attempt to ascertain whether H$_2$O or CH$_4$ is the preferred absorber in LTT~9779~b's terminator region. We conduct three additional free retrievals; one removing H$_2$O, another removing CH$_4$, and a final one removing both molecules. The prior ranges for all included parameters remain the same as stated above. Comparing the model likelihoods from each case, we are unable to determine whether the spectrum is best fit with H$_2$O ($\rm \log Z=-129.8$), CH$_4$ ($\rm \log Z=-130.2$), or a combination of both ($\rm \log Z=-129.6$) as the differences between the model evidences in each of the three cases are far from being statistically significant \citep{benneke_how_2013}. However, we find a $>$5$\sigma$ preference for the inclusion of H$_2$O and/or CH$_4$ over an atmosphere devoid of either ($\rm \log Z_{flat}=-163.4$). 

\begin{figure*} 
	\centering
	\includegraphics[width=\textwidth]{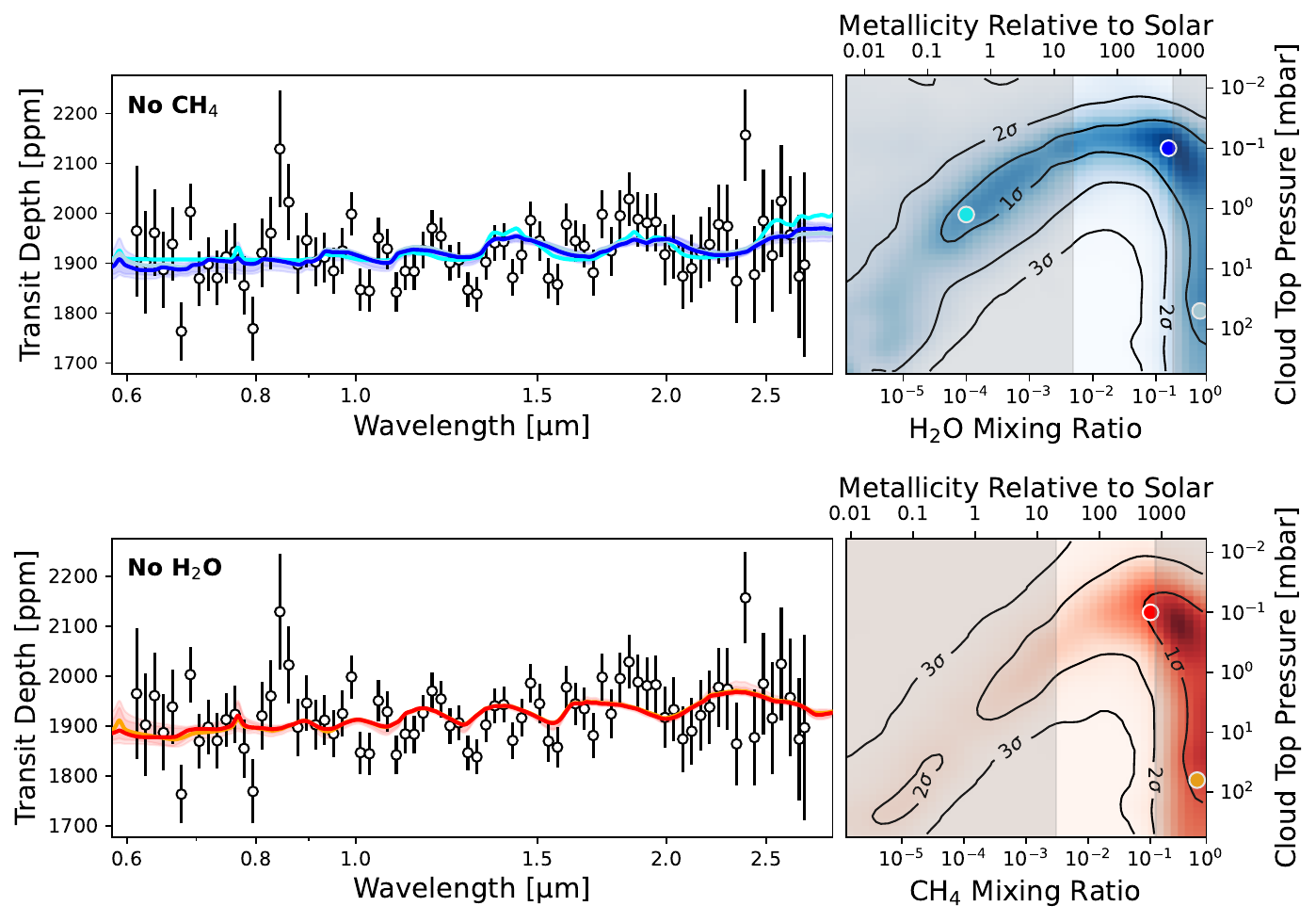}
    \caption{Water and methane dominated atmosphere models for LTT~9779~b.\emph{Top Left}: Combined transmission spectrum of LTT~9779~b. Overplotted in dark blue is the best-fitting model from our atmosphere retrievals as well as the 1 and 2$\sigma$ credible envelopes. Degeneracy between cloud pressure and metallicity \citep[e.g.,][]{benneke_how_2013, benneke_water_2019} allows for solutions with a sub-solar metallicity and deeper clouds (light-blue), as well as with extremely-high ($>$1000$\times$ solar) metallicity and clouds below the observable photosphere (blue-grey). Both data and models are binned to a constant resolution of $R=50$ for visual clarity. Note that the blue-grey model is virtually indistinguishable from the dark blue.
    \emph{Top Right}: Two-dimensional correlation between the atmosphere metallicity and cloud top pressure in our retrieved posterior distribution. The coloured points mark the parameter space ``locations" of the correspondingly coloured atmosphere model. Grey shading denotes regions of parameter space that can be ruled out by population synthesis and interior modelling work (see Section~\ref{sec: Discussion}). 
    \emph{Bottom}: Same as the top, but for the ``No H$_2$O'', CH$_4$-dominated case. Note that like the dark blue and blue-grey models above, the orange and red models are virtually indistinguishable. 
    \label{fig:spectrum}}
\end{figure*}

As shown by Coulombe et al.~(submitted), LTT 9779~b's atmosphere undergoes enormous changes in temperature across the terminator region; from $\sim$600 -- $\sim$1800\,K moving from night to dayside. It is, therefore, possible that light from the host star passing through the planet's terminator region encounters chemical regimes that are highly nonuniform as a function of longitude; with CH$_4$ dominating near the colder nightside, and H$_2$O becoming more prominent closer to the dayside. Our transmission spectrum is then effectively an average over these different regimes. We, therefore, move to consider two end-member scenarios: an atmosphere where the dominant absorbers (in the NIRISS waveband) are H$_2$O and clouds, and another with CH$_4$ and clouds. The true state of LTT 9779~b's terminator likely lies somewhere in between. Using the retrievals described in the previous paragraph, the H$_2$O/CH$_4$ vs cloud-pressure posterior distributions for each case as well as some representative atmosphere models are shown in Figure~\ref{fig:spectrum}. 

Finally, since \citet{edwards_characterizing_2023} find evidence for FeH in their HST/WFC3 spectrum, we perform another free chemistry retrieval, identical to the initial free chemistry retrieval, except also including FeH opacity \citep{Wende2010}. The abundance of FeH remains unconstrained by the data, and we are therefore not able to confirm the results of \citet{edwards_characterizing_2023}. The inference of FeH from HST likely stems from the slope seen in the transmission spectrum, which is not present in our NIRISS observations. We also test for the presence of optical absorbers such as TiO and VO, but as in the case of FeH, we find no evidence for either in the transmission spectrum.

\subsection{Interior Structure Modelling}
To ensure that our inferred atmosphere compositions are consistent with the measured bulk density of the planet, we run interior structure models in a retrieval framework following \citet{Thorngren2019}.  The models at any given time-step are solutions of the equations of hydrostatic equilibrium, mass conservation, and the material equations of state.  We use \citet{Chabrier2019} for the H/He and \citet{Thompson1990} for the metals, which we represent as a 50-50 mixture of rock and ice.  Because we are seeking an upper limit on the atmospheric metallicity, we model the conservative case (for that purpose) of a fully-mixed planet.  The models are evolved forward in time from a hot initial state using the atmosphere models of \citep{Fortney2007}.  As LTT~9799~b is a hot-Neptune, we include the anomalous hot Jupiter heating effect, parameterized in \citep{Thorngren2018}.  These models are then used in a Bayesian retrieval framework to find planet bulk metallicities consistent with the planet's mass, radius, and age (all with uncertainties). We find that LTT~9799~b is indeed a quite metal-rich planet, with $Z_p = 90 \pm 2 \%$.  
Taking the 2-$\sigma$ upper limit of our metal distribution and converting to a number ratio, we obtain a \textit{maximum} atmosphere metallicity of $\sim$850$\times$ solar.

Additionally, \citet{fortney_framework_2013} conducted a population synthesis study to attempt to predict, \textit{a-priori}, the possible atmosphere metallicities of transiting planets, including those of Neptune-like masses. They show that metallicities in excess of 1000$\times$ solar are to be expected, however, they do not find a single Neptune-mass planet with a metallicity $<$20$\times$ solar. This fits in with trends observed in both the solar system and transiting exoplanets \citep[e.g.,][]{Welbanks2019}. We can therefore set a rough lower limit for the possible metallicity of LTT~9779~b's atmosphere at $\sim$20$\times$ solar. This constraint, and that from interior structure models are shown as grey shaded regions in Figure~\ref{fig:spectrum}, and substantially restrict the possible parameter space of atmosphere compositions.

\subsection{Searching for Tracers of Atmosphere Escape}
Finally, we inspected the pixel-level (that is; one light curve per detector pixel column) transit depths near 1.083\,µm for evidence of excess absorption due to the metastable He triplet \citep{oklopcic_new_2018, spake_helium_2018}. Due to its extreme level of irradiation, LTT~9779~b receives $>$2500$\times$ more flux from its host star than the Earth, it might be expected to be in the process of losing its atmosphere through photoevaporation \citep{owen_kepler_2013, owen_photoevaporation_2018}. However, like \citet{edwards_characterizing_2023}, we find no clear evidence for He absorption in our spectrum. 

Nevertheless, we attempt to assess the strength of any potential He signal in the data, which may not be obvious by-eye, by fitting a Gaussian to the pixel-level transmission spectrum. The Gaussian He model has a fixed full-width-half-maximum of $\sim$0.75\,\AA\ (comparable to values reported by \citet{allart_spectrally_2018} and  \citet{allart_homogeneous_2023}), and centered at 1.0833\,\AA\, but with a free amplitude convolved to the resolution of NIRISS/SOSS (R$\sim$700). The fitted amplitude seen at the full resolution of SOSS is 0.15$\pm$0.61\,$\%$ \citep[as a comparison the helium signature measured for HD\,189733b is $\sim$0.7$\%$; e.g.,][]{Salz2018, Guilluy2020, allart_homogeneous_2023}.

LTT~9779, though, is a G7V star and it is thought that the UV spectra of K-type stars are more conducive to populating the metastable He triplet state \citep{oklopcic_helium_2019}. The lack of a metastable He detection therefore may not mean that LTT~9779~b is not undergoing significant mass loss, but that this mass loss cannot be probed optimally via the 1.083\,µm metastable He triplet. Other mass-loss tracers like Ly-$\alpha$ (the system is only at a distance of 80\,pc and interstellar medium absorption might not be completely optically thick), or lines of ionized metal \citep{linssen_expanding_2023} may prove to be more fruitful tracers of atmospheric escape around sun-like stars. SOSS is also potentially able to probe atmosphere loss via H$\alpha$ \citep{vidal-madjar_extended_2003, lim_atmospheric_2023, Howard2023}, though we also find no evidence for excess H$\alpha$ absorption in our data. Thus, as proposed by \citet{Fernandez2023}, due to the weak high-energy emission of the host star LTT~9779~b may simply not be undergoing sufficiently significant photoevaporative mass loss for any of the aforementioned tracers to be observable.

\section{Discussion} 
\label{sec: Discussion}

Our transmission spectrum of the unique hot-Neptune LTT~9779~b is relatively featureless and our retrievals thus prefer solutions with negligible feature amplitudes (Figure~\ref{fig:spectrum}). 
The retrievals are able to create a flat transmission spectrum with three broad families of solutions: one with very sub-solar ($<$0.1$\times$ solar) metallicity, one with super-solar metallicity and low-pressure (P$\rm _{cloud}\lesssim1$\,mbar) clouds, and a final family with extremely high ($>$500$\times$ solar) metallicity (Figure~\ref{fig:spectrum}). This is a manifestation of the well-known cloud-metallicity degeneracy \citep[e.g.,][]{benneke_how_2013,Knutson2014,benneke_water_2019}: a low column density of H$_2$O or CH$_4$ vapour is the primary cause of the lack of features in the low metallicity case, whereas the feature amplitudes in the second family of models are dampened by clouds. In the final model family, the atmosphere metallicity is sufficiently elevated to reduce the atmosphere scale height and attenuate the amplitude of molecular features without needing to invoke clouds.

Due to the flatness of LTT~9779~b's transmission spectrum, we are not able to make any strong claims about the atmosphere composition from the transit observations alone. Previous studies of canonical Neptune-mass planets such as GJ~436~b have demonstrated that exotic atmosphere compositions, with metallicities in excess of 1000$\times$ solar, are indeed possible \citep{fortney_framework_2013, moses_compositional_2013, morley_forward_2017}. However, we can begin to limit the parameter space of possible atmosphere compositions through arguments of self-consistency. In particular, our interior structure upper limit rules out high-metallicity/cloud-free atmospheres at $\sim$2$\sigma$, and population synthesis constraints remove the low-metallicity scenarios; irrespective of whether H$_2$O or CH$_4$ is assumed to be the dominant absorber.  The most likely family of solutions thus becomes that with elevated metallicity between $\sim$20 and 850$\times$ solar, and clouds at $\sim$mbar pressure levels --- though atmospheres with metallicities $>$500$\times$ solar and cloud-free photospheres cannot be entirely ruled out. 

An atmosphere with an elevated metallicity and clouds is consistent with findings from previous studies of this planet. In particular \citet{hoyer_extremely_2023} and Coulombe et al.~(submitted) both find evidence for a high dayside albedo ($A_g$ = 0.6 -- 0.8). As shown in both of those works, the presence of reflective, high-altitude clouds can account for this elevated albedo. The cloud-free low or high metallicity scenarios allowed by our data would thus not be consistent with previous findings. In theory, exotic cloud species could emerge at the extreme high-metallicity end of our probed parameter space. \citet{moses_compositional_2013}, in particular, explore the condensation of graphite in $>$1000$\times$ solar, carbon-rich atmospheres of Neptune-mass planets, and conclude that it is unlikely to build up in the photosphere in sufficient concentrations to affect transmission spectra.

\begin{figure} 
	\centering
	\includegraphics[width=\columnwidth]{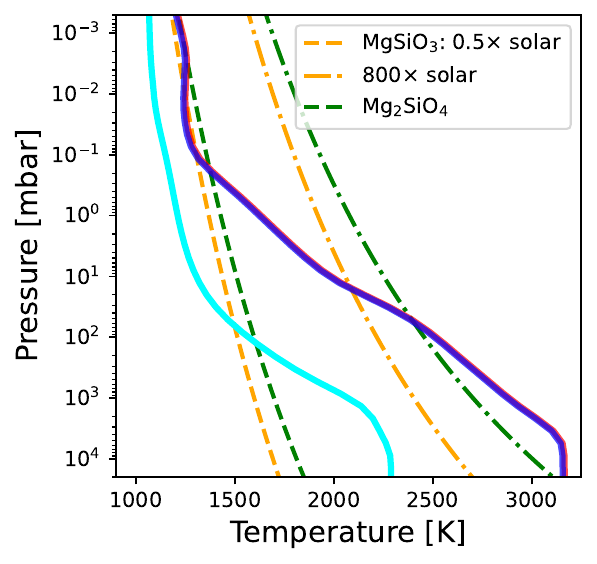}
    \caption{Exploration of cloud formation in LTT~9779~b's terminator region via a comparison of terminator PT profiles and condensation curves of prominent silicate clouds species. PT profiles were calculated self-consistently using the SCARLET framework \citep{benneke_strict_2015}, for three of the cases shown in Figure~\ref{fig:spectrum}: sub-solar metallicity H$_2$O-dominated (light blue), super-solar metallicity H$_2$O-dominated (dark blue), and super-solar metallicity CH$_4$-dominated (red). Condensation curves are from \citet{visscher_atmospheric_2010} and calculated at two metallicities: 0.5$\times$ solar (dashed), 800$\times$ solar (dot-dashed), reflecting the sub- and super-solar metallicity models shown. The terminator temperature of LTT~9779~b is sufficiently cool to allow for silicate clouds to readily condense at millibar pressures, even for sub-solar atmospheric metallicities. Note that the red and dark blue PT profiles are entirely overlapping. 
    \label{fig:PT}}
\end{figure}

Based on the above considerations, we explore the thermochemistry of possible cloud condensation with our retrieved terminator conditions. Using the SCARLET framework \citep{benneke_strict_2015}, we generate self-consistent atmosphere pressure-temperature (PT) profiles for metallicities of $\sim$0.5$\times$ and $\sim$800$\times$ solar, corresponding to the light blue and dark blue/red models in Figure~\ref{fig:spectrum}, respectively. The models were computed considering a bond albedo of $A_B$ = 0.66, heat redistribution of $f$ = 0.5 (uniform dayside), and no opacity from optical absorbers such as TiO and VO (which would lead to a temperature inversion) as informed from the thermal emission retrievals shown in Coulombe et al.~(submitted). We show these PT profiles, as well as condensation curves for enstatite (MgSiO$_3$) and forsterite (Mg$_2$SiO$_4$) from \citet{visscher_atmospheric_2010} in Figure~\ref{fig:PT}, demonstrating that silicate clouds can readily condense at mbar levels in LTT~9779~b's terminator region, even at solar atmospheric metallicities. Advection of these clouds onto the day-side could then cause the high albedo reported by \citet{hoyer_extremely_2023}.

Though the low XUV output of the host star likely renders any photoevaporative mass loss processes very inefficient \citep[e.g.,][]{Fernandez2023}, a core-powered mass loss mechanism (that is, mass loss driven by the host star's bolometric luminosity as opposed to just its high-energy output; e.g., \citealp{Ginzburg2018, Gupta2021}) may still be impactful. We posit that a cloudy nature for LTT~9779~b's atmosphere may be part of a positive feedback loop that suppresses the effectiveness of atmospheric loss and contributes to its survival in the desert. For instance, following \citet{seager_textbook}, the equilibrium temperature of a planet can be calculated;
\begin{equation}
    T_\mathrm{eq} = T_\mathrm{eff,*}\bigg(\frac{R_*}{a}\bigg)^{1/2}[f(1-A_\mathrm{B})]^{1/4}
\end{equation}
where $T_\mathrm{eff,*}$ is the host star effective temperature, $A_\mathrm{B}$ the planet's bond albedo, and $f$ a factor describing heat recirculation. Assuming partial heat recirculation \citep{hoyer_extremely_2023} and the scaled semi-major axis and host star temperature from \citet{jenkins_ultra-hot_2020}, increasing the bond albedo from 0 to $\sim$0.7 (consistent with \citet{hoyer_extremely_2023} and Coulombe et al.~(submitted)), results in a decrease in $T_\mathrm{eq}$ from $\sim$2300\,K to $\sim$1600\,K. Again, following \citet{seager_textbook}, the thermal escape parameter, $\lambda_c$, is;
\begin{equation}
    \lambda_c = \frac{m_\mathrm{H}g_\mathrm{e}r_\mathrm{e}}{k_\mathrm{B}T_\mathrm{e}}
\end{equation}
where $g_\mathrm{e}$, $r_\mathrm{e}$, and $T_\mathrm{e}$ are the gravity, radius, and temperature respectively of the exobase, and $m_\mathrm{H}$ the mass of hydrogen. Smaller escape parameters correspond to larger escape fluxes via $F_{esc}\propto(1+\lambda_c)e^{-\lambda_c}$. Assuming that the exobase temperature is proportional to the equilibrium temperature results in an increase in the escape parameter by a factor of $\sim$1.5 for the $A_\mathrm{B}=0.7$ compared to the $A_\mathrm{B}=0$ case. Though this is not enough to completely halt any escape processes, or change the escape regime from hydrodynamical to hydrostatic \citep[e.g.,][]{pierrehumbert_textbook}, it will, nevertheless, decrease the rate at which the atmosphere is escaping. Supposing LTT~9779~b initially formed as a cloud-free hot-Saturn \citep[e.g.,][]{jenkins_ultra-hot_2020, Fernandez2023}, the preferential loss of light elements (i.e., hydrogen) over time could result in an increase in atmosphere metallicity, or decrease in gravity; both of which favour cloud formation \citep[e.g.,][]{stevenson2016, hoyer_extremely_2023}. Clouds then cause a decrease in the efficiency of atmosphere loss, prolonging the lifetime of the atmosphere.

\section{Conclusions} 
\label{sec: Conclusions}

In this work, we have presented the first JWST transmission spectrum of LTT~9779~b; the only known hot-Neptune to have retained a substantial atmosphere. In light of population synthesis and interior modelling considerations, cloudy, high-metallicity solutions are the most likely explanations for our transmission spectrum, regardless of whether we consider a H$_2$O or CH$_4$-dominated atmosphere --- though our transit data alone also allows for sub-solar and $\sim$1000$\times$ solar metallicity solutions with cloud-free photospheres. Along with Coulombe et al.~(submitted), this work lays the groundwork for an understanding of the intriguing hot-Neptune LTT~9779~b. Coupled with the spectrum presented here, future observations of this planet with JWST NIRSpec\footnote{GO 3231; PI Crossfield} should have the necessary precision to differentiate between H$_2$O and CH$_4$-dominated terminator conditions, and better refine the atmosphere metallicity.

\begin{acknowledgments}
M.R.\ would like to acknowledge funding from the Natural Sciences and Research Council of Canada (NSERC), as well as from the Fonds de Recherche du Quebec Nature et Technologies (FRQNT). He would also like to thank David Grant and Lili Alderson for useful and entertaining conversations about blue-ish noise.
L.-P.C.\ acknowledges funding by the Technologies for Exo-Planetary Science (TEPS) Natural Sciences and the NSERC CREATE Trainee Program.
J.T.\ is supported by the Eric and Wendy Schmidt AI in Science Postdoctoral Fellowship, a Schmidt Futures program. 
R.A.\ is a Trottier Postdoctoral Fellow and acknowledges support from the Trottier Family Foundation.
D.J.\ is supported by NRC Canada and by an NSERC Discovery Grant.
This project was undertaken with the financial support of the Canadian Space Agency.
L.D.\ acknowledges support from the Banting Postdoctoral Fellowship
program, administered by the Government of Canada.
This work is based on observations made with the NASA/ESA/CSA JWST. The data were obtained from the Mikulski Archive for Space Telescopes at the Space Telescope Science Institute, which is operated by the Association of Universities for Research in Astronomy, Inc., under NASA contract NAS 5-03127 for JWST. 
The specific observations analyzed can be accessed via~\dataset[10.17909/8mg4-c402]{10.17909/8mg4-c402}.
This research has made use of the NASA Exoplanet Archive, which is operated by the California Institute of Technology, under contract with the National Aeronautics and Space Administration under the Exoplanet Exploration Program.
\end{acknowledgments}

\vspace{5mm}
\facilities{JWST(NIRISS), Exoplanet Archive}

\software{\texttt{astropy} \citep{astropy:2013, astropy:2018}, \texttt{celerite} \citep{foreman-mackey_fast_2017},
\texttt{ExoTiC-LD} \citep{david_grant_2022_7437681},
\texttt{ipython} \citep{PER-GRA:2007},
\texttt{juliet} \citep{espinoza_juliet_2019},
\texttt{jwst} \citep{bushouse_howard_2022_7038885},
\texttt{matplotlib} \citep{Hunter:2007},
\texttt{numpy} \citep{harris2020array},
\texttt{pymultinest} \citep{buchner_statistical_2016},
\texttt{scipy} \citep{2020SciPy-NMeth}
}

\appendix

\begin{figure*} 
	\centering
	\includegraphics[width=\textwidth]{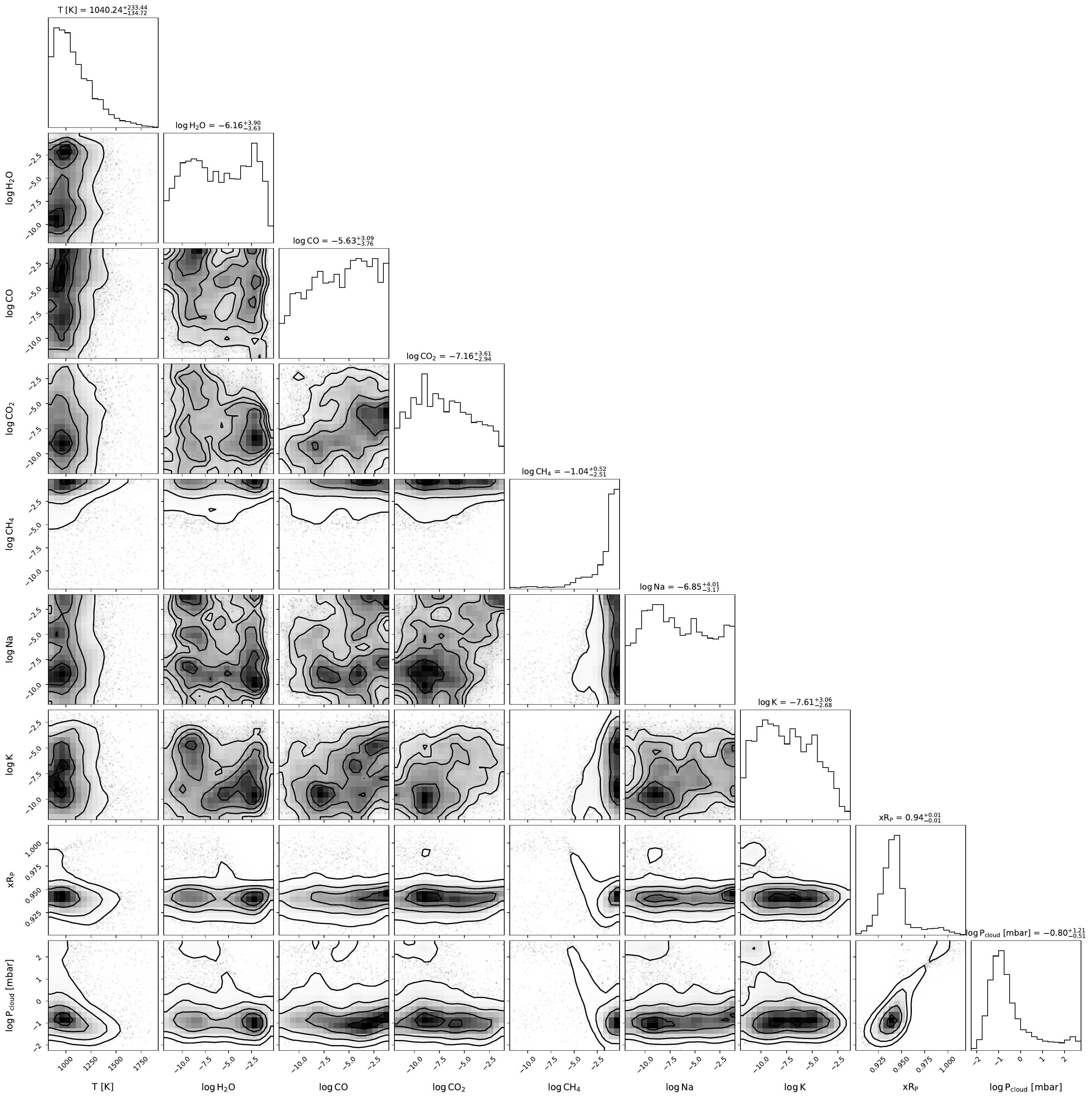}
    \caption{Posterior distributions for the full free chemistry retrieval described in Section~\ref{sec: Modelling}. There are weak preferences for the presence of H$_2$O, CH$_4$, and clouds at sub-mbar pressures. In many cases, the ``best-fit'' values reported above the marginal distributions are simply the mean of the prior range. 
    \label{fig:corner}}
\end{figure*}

\bibliography{main}{}
\bibliographystyle{aasjournal}

\end{document}